\documentclass[reprint, amsmath,amssymb,aps,prb]{revtex4-1}

\usepackage{textcomp}%Provides extra symbols, e.g. arrows like \textrightarrow, various currencies (\texteuro,...), things like \textcelsius and many others.
\usepackage{graphicx}% Include figure files
\usepackage[outdir=./]{epstopdf}%Provides and option to convert EPS images to PDF and include them with \includegraphics{}.
\usepackage{dcolumn}% Align table columns on decimal point
\usepackage{bm}% bold math
\usepackage{tabularx} % Ruled Tabular
\usepackage{lipsum}
\usepackage{multirow}
\usepackage{times}
\usepackage{wrapfig}
\usepackage{xcolor}
\usepackage{subcaption}
\usepackage{listings}
\usepackage{silence}
\usepackage[maxfloats=1000]{morefloats}
\def\bmg{$Pd_{40}Ni_{40}P_{20} $\hspace{0.05cm}}
\def\our{Pd-Ni-P\hspace{0.01cm}}

\begin{document}

\preprint{APS/123-QED}

\title{ \textit{Ab initio}  inversion of structure and the lattice dynamics of a metallic glass: The case of {\bmg}}
\author{Bishal Bhattarai}
\email{bbhattarai@wustl.edu}
\affiliation{Department of Mechanical Engineering and Materials Science, \\ Washington University in St. Louis, St. Louis, Missouri 63130, United States}

\author{Rajendra Thapa}
\email{rt887917@ohio.edu}

 \author{D. A. Drabold }%
\email{drabold@ohio.edu}
\affiliation{Department of Physics and Astronomy \\ 
 Ohio University, Athens, Ohio 45701, United States}

\date{\today}

\begin{abstract}
In this paper we infer the structure of \bmg from experimental diffraction data and {\it ab initio} interactions using "Force Enhanced Atomic Refinement" (FEAR). Our model accurately reproduces known experimental signatures of the system and is more efficient than conventional melt-quench schemes.   We critically evaluate the local order, carry out detailed comparisons to extended X-ray absorption fine structure (EXAFS) experiments and also discuss the electronic structure. We thoroughly explore the lattice dynamics of the system, and describe a vibrational localized-to-extended transition and discuss the special role of P dynamics. At low energies P is fully contributing to extended modes, but at higher frequencies executes local motion reminiscent of a "rattler" inside a cage of metal atoms.  These highly localized vibrational states suggest a possible utility of these materials for thermoelectric applications.
\end{abstract}

\pacs{Valid PACS appear here}
\maketitle

\section{\label{sec:level1}INTRODUCTION }

Bulk metallic glasses (BMG) are materials formed with cooling rates faster than 10\textsuperscript{3} K/s or one which has 
a thickness greater than 1mm~\cite{Inoue}, and are fundamentally different from traditional amorphous alloys formed at a very high cooling rates to suppress nucleation of crystalline phases~\cite{Perim,Kelton2}. They are amorphous alloys that exhibit a glass transition at some characteristic temperature. They exhibit extreme strength at low temperatures, and high flexibility that enables the use of BMG as soft tissue stents, providing improved compliance with blood vessel biomechanics and minimal damage to vessels~\cite{Schroers}. They show an abrupt change in thermodynamic and physical properties at the glass transition temperature ($T_g$)~\cite{Chen}. 
After the initial discovery of the materials in 1959 at Caltech ~\cite{Klement}, BMGs gained a lot of attention and at present provide fundamental scientific puzzles and diverse applications ranging from sporting goods to micro electromechanical systems (MEMS), nanotechnology to biomedical applications~\cite{Biomed,EvanMa}. The diverse applicability arises from the properties like high strength, resistance
to wear and corrosion~\cite{Kumar}, etc. which are attributed to the amorphous state, possessing no dislocations or grain boundaries.

The structure of BMG is controversial, particularly for metal-metalloid-based BMG such as
 {\our} -- the structure is complex with diverse interpretations appearing in the literature~\cite{Schwarz,Kimura,Inoue1,Matsubara,Gaskell}. One of the earliest models, Bernal's dense random packing model,~\cite{Bernal,Bernal1} satisfactorily explains monoatomic metals but fails to provide structural models for 
multi-component glassy systems and metal-metalloid-based alloys with pronounced chemical short-range order.  Another stereo-chemically defined model of Gaskell assumes that the local units of nearest neighbors in amorphous
metal-metalloid-based alloys should have the same type of structure as the corresponding crystalline compound with similar configuration~\cite{Gaskell,Gaskell1,Gaskell2}. Recently, researchers have also devised a hybrid atomic
packing scheme in metal-metalloid-based glasses~\cite{Guan}. Another model for BMGs is the dense packing of atomic clusters developed by D.B.Miracle~\cite{Miracle2,Miracle1}.

Among metallic glasses, {\bmg}  is popular for several reasons:  relatively low cooling rates, (relatively) simple ternary composition, excellent glass forming ability, etc~\cite{Guan}. Using experimental probes: extended X-ray absorption fine structure (EXAFS) and extended electron energy loss fine structure (EXELFS), \textit{Alamgir et. al.}~\cite{Alamgir} reported that {\bmg} is the best glass former for the $Pd_{x}Ni_{80-x}P_{20}$ stoichiometry. Similarly, theoretical based study have also highlighted excellent glass forming ability of {\bmg} glass~\cite{Kumar,Guan}. Very recently, it was also reported that {\bmg} glasses near this composition exhibit polyamorphism and anomalous thermodynamics~\cite{Lan2017}.

A detailed atomic analysis is required to gain insight,  because the system lacks long-range order or periodicity. This system is scientifically important, its structure is unclear, and it is an ideal system to study using chemically accurate methods. To provide accurate computer models,  we have used two different methods: (a) we create a molecular dynamics (MD) based model using ``\textit{melt and quench}'' (MQ) technique where a thermally equilibrated liquid is quenched using dissipative dynamics, (b) we also prepare another model by systematically combining experimental and theoretical information.

The first approach \textit{``melt and quench"} (MQ)  is the canonical method to study amorphous materials~\cite{Medelev1}. The MD based approach produces reasonable structure when ordering in the system is quite local and structure of liquid is essentially similar to quenched glass~\cite{DraboldEuro1}. MQ based models using \textit{ab initio} method are computationally expensive and are restricted by system size $\sim$ 100-200 atoms. The availability,  accuracy and transferability of empirical interatomic potential has limited modeling of BMG in few compositions~\cite{Sha1}. 

\begin{figure}[htp]
  \centering
   \includegraphics[width=0.90\linewidth]{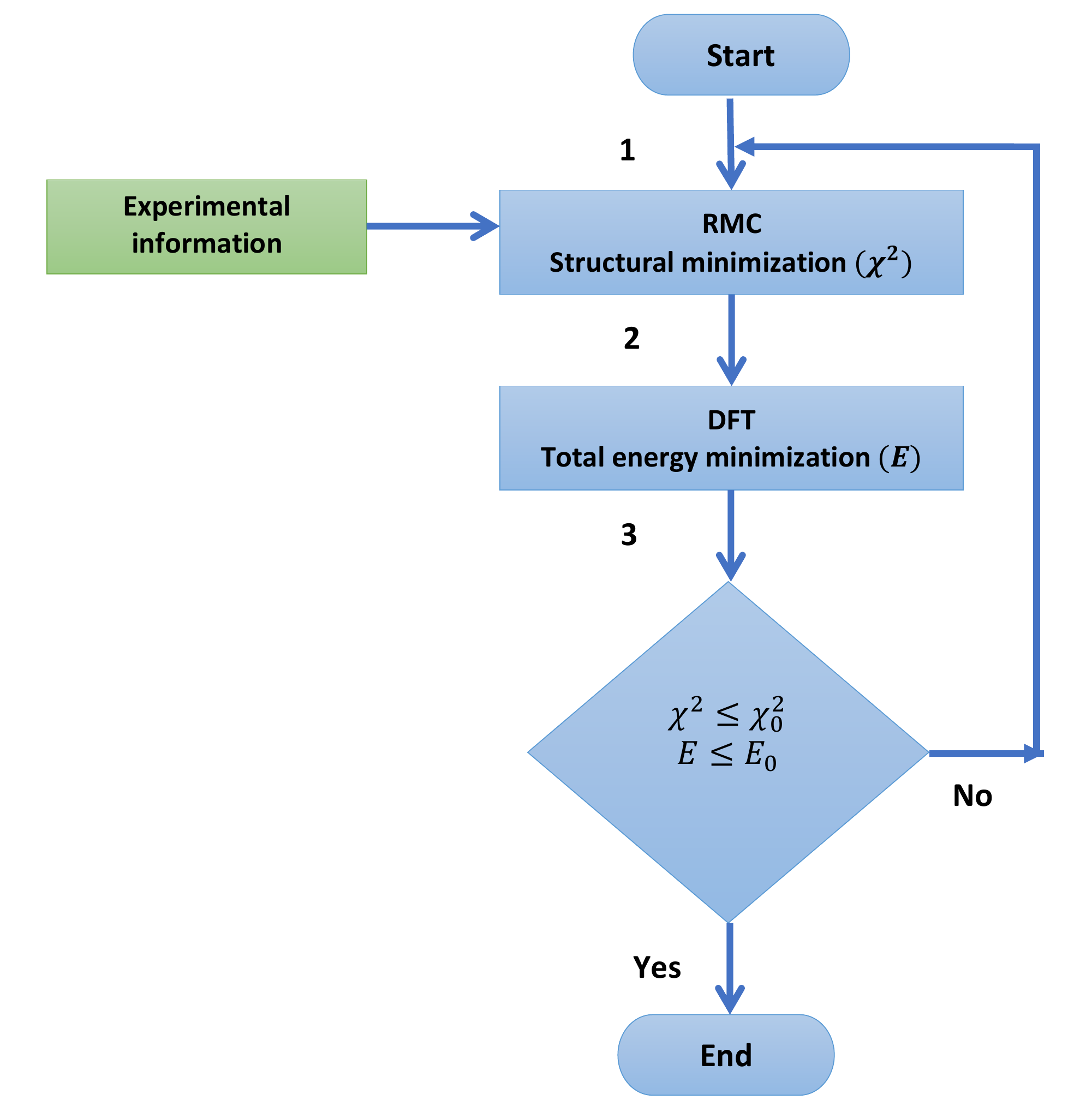}
  \caption{\textbf{Flowchart of AIFEAR method:} In AIFEAR, we start with \textit{randomly} chosen coordinates.             \textbf{(Step 1)} In the first step  randomly chosen coordinates are subjected to partial structural refinement consistent with experimental information for ``$\mathcal{N}$" accepted RMC steps. \textbf{(Step 2)} After partial structural refinement we perform partial relaxations using conjugate gradient (CG) scheme for ``$\mathcal{M}$" steps in VASP. \textbf{(Step 3)}  The process of 
  partial structural refinement and partial relaxation is repeated until the model is fully converged to defined accuracy. The final results do not heavily depend upon the values of $\mathcal{N}/ \mathcal{M}$. A ratio of $50/1$ was used for $\mathcal{N}/ \mathcal{M}$ with 5 CG steps per FEAR step.}\label{Fig:FearFlowchart}
\end{figure}

Our second approach uses a novel \textit{ab initio} based structural inversion method: force enhanced atomic refinement (AIFEAR)~\cite{Anup,Anup2,Anup3,Bishal1,Bishal2,Bishal3}. Structural inversion of complex metallic glass has been a useful tool to provide insights to the material properties~\cite{Hwang,Sheng}.  The need to incorporate {\it a priori} experimental information is almost obvious, but a conventional Reverse Monte Carlo (RMC)~\cite{McGreevy} approach  produces incorrect chemical ordering and an otherwise overly disordered model unless one includes additional constraints to compel specified local order. On the other hand, incorporating multiple constraints in cost function makes inversion problem more challenging and of course biases the resulting model. To overcome this hurdle and make effective use of experimental information available, several hybrid methods~\cite{Opletal5,ECMRGeSe2} have been developed. The \textit{ab initio} based force enhanced atomic refinement (AIFEAR) is one such approach. AIFEAR has thus far proven to be a robust  and unbiased method to model diverse amorphous materials. AIFEAR is an iterative means to invert diffraction data while simultaneously finding  appropriate coordinates minimizing \textit{ab initio} forces and energies (see Fig.~\ref{Fig:FearFlowchart}).  AIFEAR has an obvious advantage over usual MQ approach as it is significantly less computationally expensive compared to the MQ approach (requiring fewer force calls compared to typical MQ models), thus enabling us to prepare large realistic models (for example, 1024 atoms for a-Si\cite{Bishal2} and 800 atoms for a-graphene~\cite{Bishal3}).  The details of FEAR approach has been presented elsewhere~\cite{Anup2,Anup3,Bishal1}. 

We carry out a thorough study of the vibrational properties.  We track the character of the phonons as a function of frequency across the entire spectrum and elucidate the character of a localized-delocalized transition in the range of  250-400 cm $^{-1}$.  To properly represent lower energy modes, we have used two \textit{ab initio} based models of size 200  and 300 atoms with plane wave basis set with a reasonable plane wave cut off. We show that there is an interesting, and apparently continuous localized-to-extended transition in the normal modes, which we illustrate and explain. This transition would have a significant impact on thermal transport in the materials.

The rest of the paper is organized as follows: In section II we discuss details about computational methodology and model generation. In Section III we present our results of structural, electronic and vibrational properties by comparing with experiment and previous literature. Section IV summarizes our findings and important discussions.

\section{Methodology and Models}

\subsection{Model I: MQ200}

We prepare a MQ model of 200 atoms consistent with the experimental density 9.40 g/cc~\cite{Haruyama}, starting with random coordinates, using the \textit{ab initio} plane-wave density functional theory (DFT) package VASP~\cite{vasp,vasp1,vasp2}. Our calculation is carried out with the projector augmented wave (PAW)~\cite{PAW} method with a generalized gradient approximation~\cite{PBE} for the exchange-correlation potential. The model is first ``heated" well above melting temperature to form a liquid at 3000 K. The model is then equilibrated at 3000 K for another 8 ps  to remove any possible bias. Finally this well equilibrated liquid is arrested into a glassy structure by cooling and equilibrating in multiple steps at 2000 K, 1000 K and 300 K. The molecular dynamics (MD) simulation is performed using a time step of 2.0 fs with a total simulation time of 47 ps. Our simulations are performed with a single k-point $\Gamma (k = 0)$. 

\subsection{ Model II: FEAR300}

In this section we present the details of our AIFEAR model. A flowchart of AIFEAR is shown in Fig.\ref{Fig:FearFlowchart}. In this approach we prepare a model of 300 atoms at same experimental density of  9.40 g/cc. We being with randomly chosen coordinates with every atom satisfying a minimum approach distance of 2.00 $\mathring{A}$  (no two atoms are allowed to be on top of each other). We structurally refine these coordinates with the RMC code RMCProfile~\cite{RMCProfile} using experimental diffraction data~\cite{Hruszkewycz}. A RMC step size of 0.085 $\mathring{A}$, a minimum approach of 2.00 $\mathring{A}$ and 0.070 weight of the experimental data was chosen for the structural refinement. In the relaxation step we use the DFT code VASP to relax the system partially.  Our AIFEAR model required 725 FEAR steps, a total of 3625 force calls to converge  compared to 23,500  force calls for the MQ model. This is $\sim 16$ \% of the total time taken by the MQ model. The final set of coordinates are relaxed using same VASP conditions as Model I i.e. a plane-wave basis set, plane-wave cutoff of 550 eV and an energy convergence tolerance of $10^{-5}$ eV and $\Gamma (k=0)$ .

\begin{table}[htp]
\setlength{\arrayrulewidth}{0.60mm}
\centering
\begin{tabular}{|c |c |c |c |c |}
\hline
Atom & $n$ & $n(Pd)$ & $n(Ni)$ & $n(P)$ \\ \hline
\rule{0pt}{4mm}

$Pd$ & 13.175 & 6.62 & 4.65  & 1.905   \\
\rule{0pt}{4 mm}

$Ni$ & 12.15 & 4.65 & 5.3 & 2.2 \\
\rule{0pt}{4 mm} 

$P$ & 8.95 & 3.82 & 4.4 & 0.73 \\
\hline 
\end{tabular}%\label{Fear:tablea}
\caption{\textbf{Total coordination statistics for FEAR300:} Average coordination number $(n)$ and its distribution onto constituent atoms of FEAR300 model. Both Pd/Ni atoms tend to mostly form bonds among themselves. }
\end{table}

\begin{table*}[htp]
\setlength{\arrayrulewidth}{0.60mm}
\centering
\begin{tabular}{|ccccc|ccccccc|ccccc|}
\hline 
\rule{0pt}{4mm}
& & $Pd$ & & && & & $Ni$& && & & & $P$& & \\
\hline \hline
\rule{0pt}{4 mm}
$n$ & ${f}$ & $Pd$ & $Ni$  & $P$ & &$n$ & $f$ & $Pd$  & $Ni$  & $P$ & & $n$ & ${f}$ & $Pd$  & $Ni$  & $P$  \\ \hline
\rule{0pt}{5 mm}

11 & 5 & 65.45 & 21.82  & 12.73  & &  10 & 5 & 53.32 & 31.67 & 15 & & 6 & 1.67 & 83.33 & 16.67 & 0 \\
\rule{0pt}{5 mm}

12 & 22.5 & 62.5 & 25.6 & 11.90 & & 11 & 20 & 45.45 & 36.36  & 18.19 & & 7 & 5 & 71.42 & 28.58 & 0\\
\rule{0pt}{5 mm} 

13 & 34.17 & 52.80 & 32.16 & 15.04 & & 12 & 43.33 & 39.85 & 45.29 & 14.86 & & 8 & 25 & 52.78 & 40.28 & 6.94\\
\rule{0pt}{5 mm} 

14 & 26.66 & 40.63 & 44.20 & 15.17 & & 13 & 21.67 & 33.48 & 46.61 & 19.91 & & 9 & 38.33 & 43.21 & 50.61 & 6.18\\
\rule{0pt}{5 mm} 

15 & 10 & 31.85 & 51.85 & 16.30 & & 14 & 6.66 & 33.33 & 50 & 16.67 & & 10 & 25 & 30 & 56 & 14 \\
\rule{0pt}{5 mm} 

16 & 1.67 & 40.63 & 43.75 & 15.62 & & 15 & 3.34 & 26.67 & 55.55 & 17.78 & & 11 & 5 & 15.15 & 63.63 & 21.22\\
\hline 
\rule{0pt}{5 mm}
$total $& 100 & 49.24 & 36.27 & 14.49 & & & 100 & 38.90 & 44.20 & 16.90 & & & 100 & 41.49 & 49.36 & 9.15\\
\hline 
\end{tabular}
\caption{\textbf{Coordination statistics for FEAR300:} Distribution of coordination number ($n$) of constituent atoms among their neighbors for FEAR300.
$f$ denotes the percentage of $n$-fold coordinated atom for the atom-species being considered. The atomic symbol($Pd,Ni,P$) on top of each table denotes the atom species being considered. All quantities, except $n$, are expressed as a percentage. The fraction of $Pd$ atom coordination is highest for $n$ being 13 and 14 and mostly tends to bond with other $Pd$ atoms. Similarly, $Ni$ atoms tend to form 12 fold-coordinated structures. The $P$ atom has slightly fewer neighbors with 9-fold coordinated atoms being the most common.}\label{Fear:tableb}
\end {table*}

\begin{figure*}[htp]
\begin{minipage}[b]{.475\textwidth}
  \centering
  \includegraphics[width=0.98\linewidth]{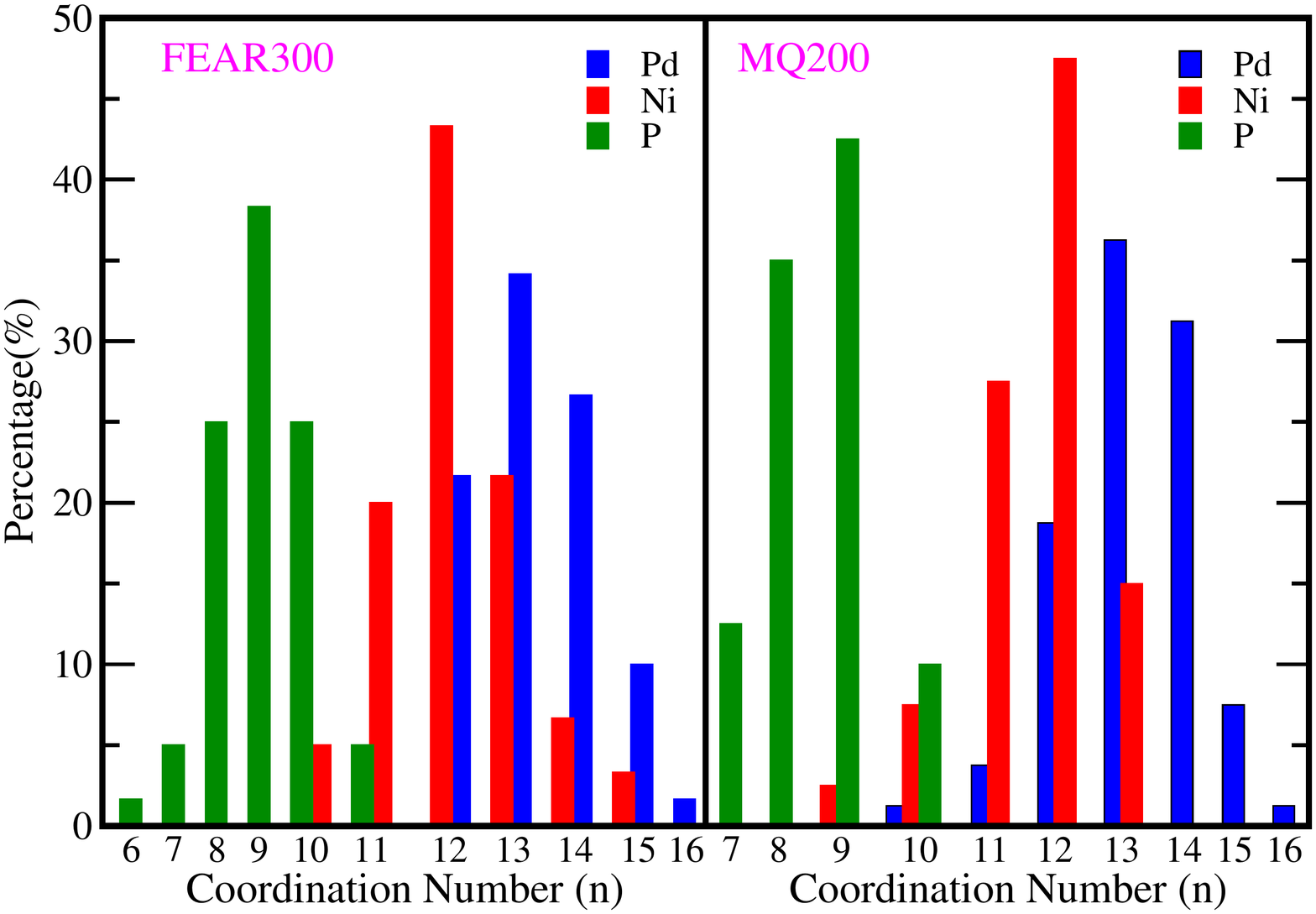}
\end{minipage}\hspace{0.1cm}
\begin{minipage}[b]{.475\textwidth}
 \centering
  \includegraphics[width=0.93\linewidth]{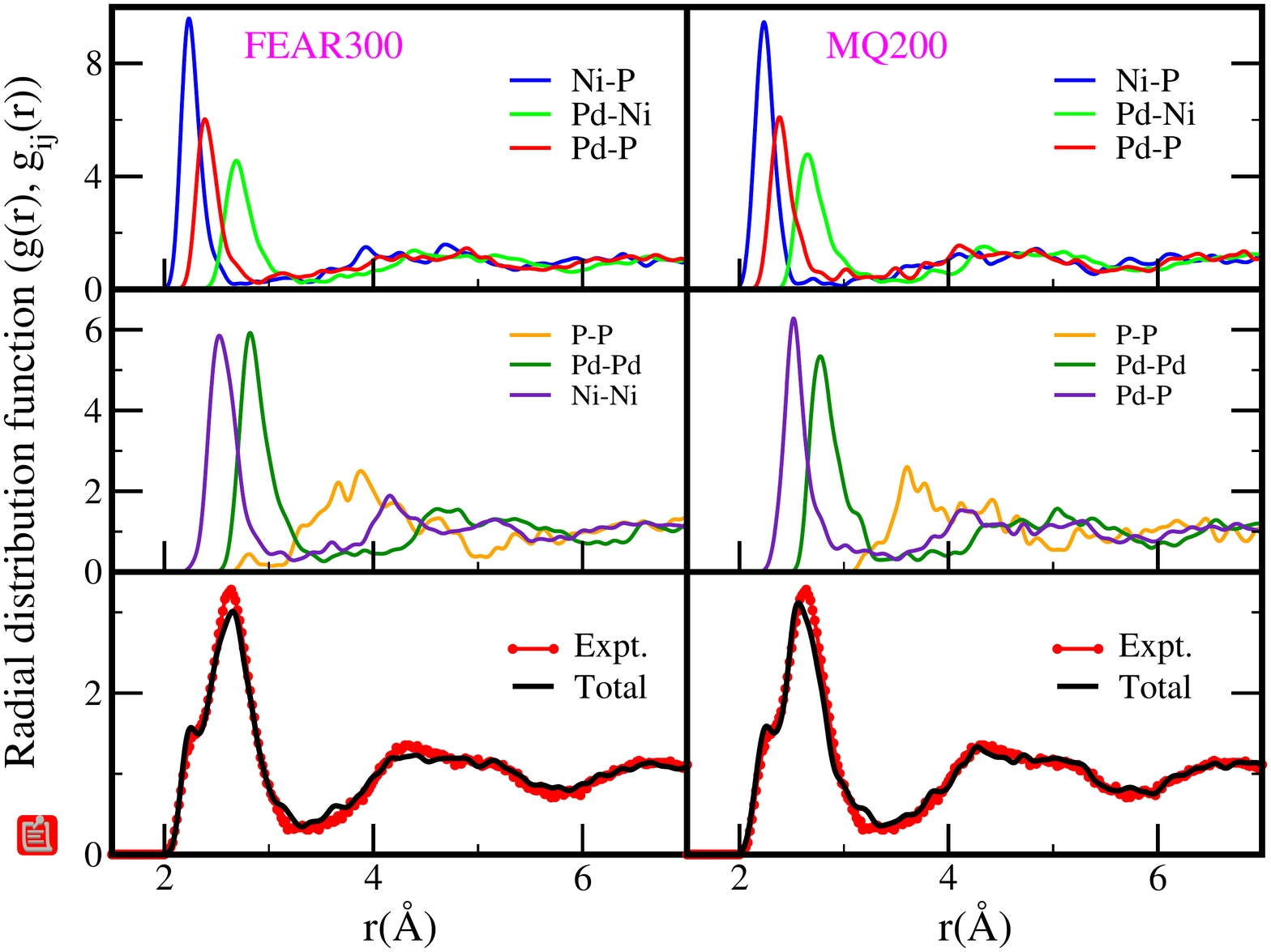}
\end{minipage}\vspace{0.1cm}
   \caption{\textbf{Coordination statistics and radial distribution function:} (\textbf{Left panel}) Coordination distribution of FEAR300 and MQ200. Both models show strikingly similar neighbor environments. The maximum number of neighbors around an atomic species is observed directly related to its atomic size. (\textbf{Right panel}) We show comparison of our models with the experiment~\cite{Hruszkewycz}. The radial distribution function ($g(r)$) is in good agreement with the experiment and the partial RDF are also consinstent with previous literatures~\cite{Kumar,Guan}.}\label{Fig_Coordination}
\end{figure*}

\begin{figure*}[!ht]
\begin{minipage}[b]{.46\textwidth}
  \centering
  \includegraphics[width=0.95\linewidth]{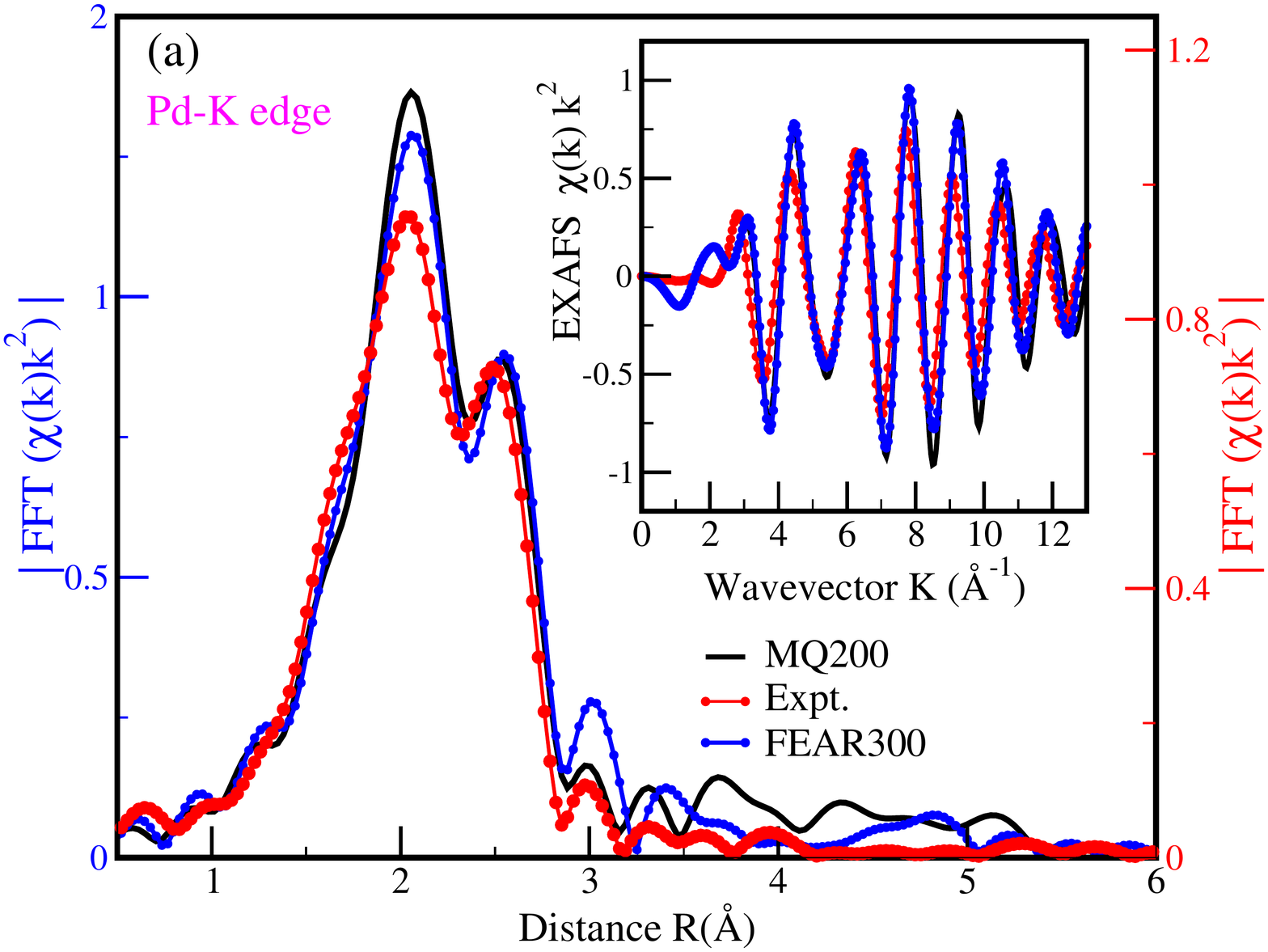}
\end{minipage} \vspace{0.1cm}
\begin{minipage}[b]{.46\textwidth}
  \centering
  \includegraphics[width=0.95\linewidth]{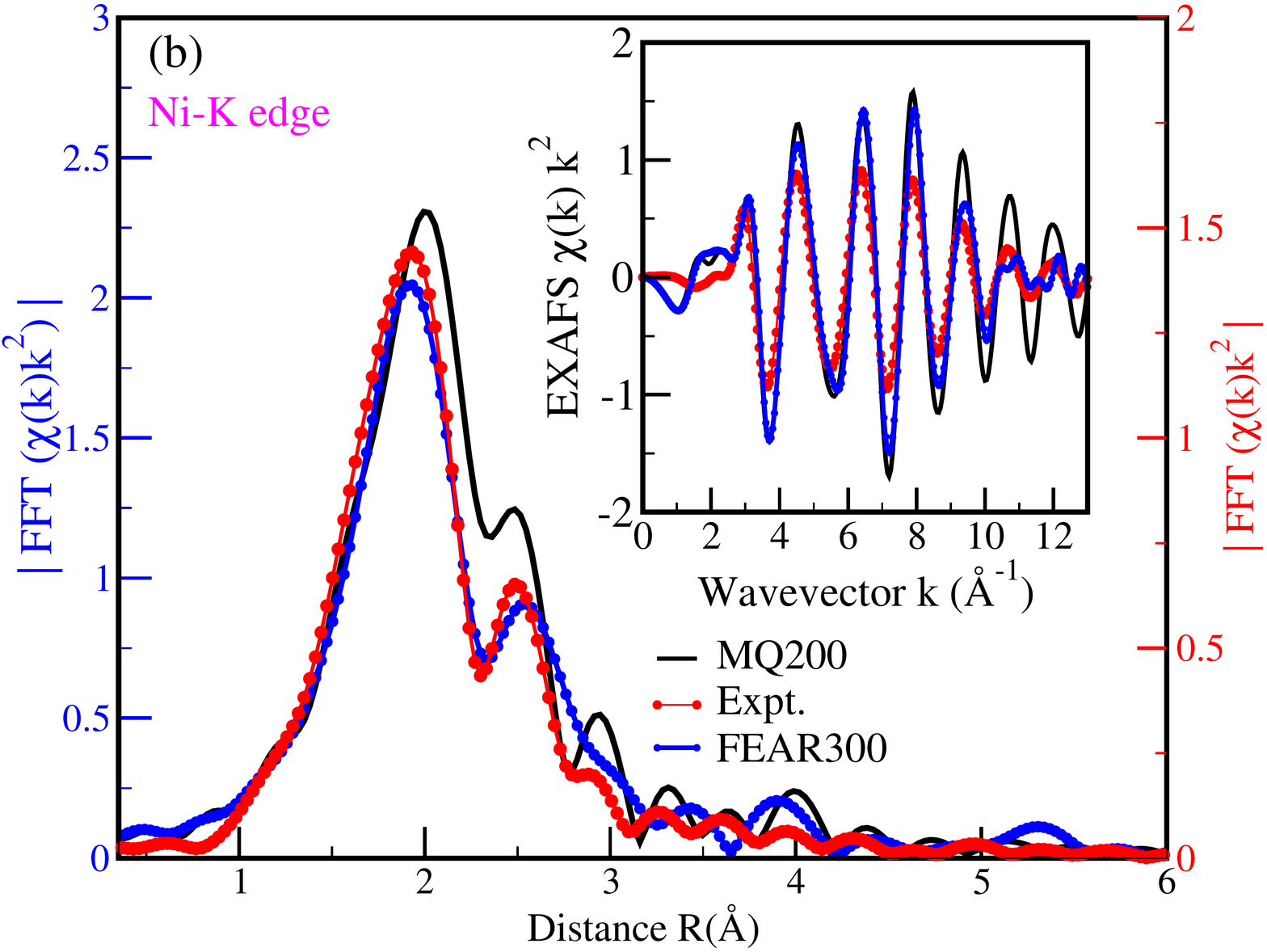}
\end{minipage}\vspace{0.1cm}
\begin{minipage}[b]{.46\textwidth}
 \centering
  \includegraphics[width=0.95\linewidth]{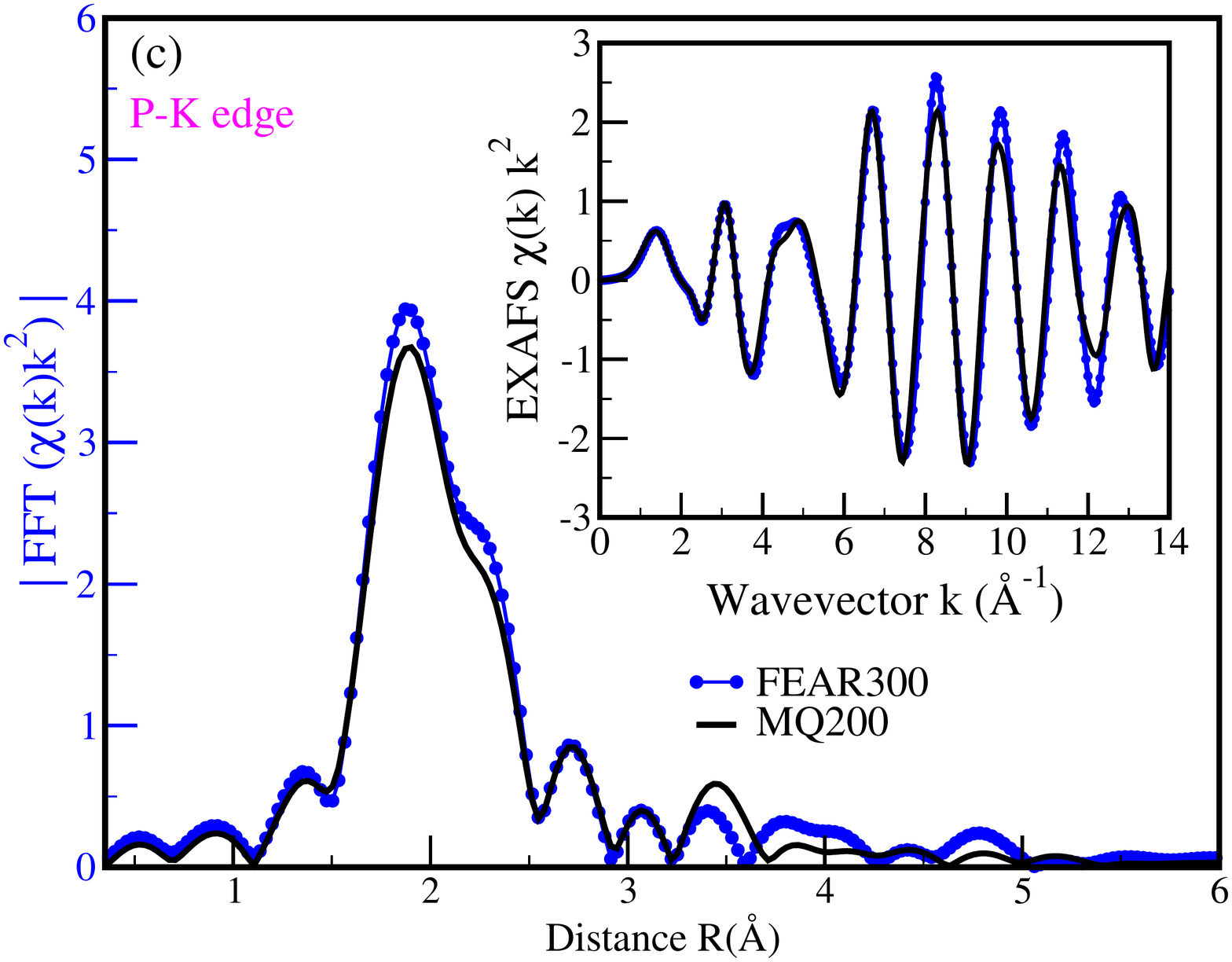}
\end{minipage}\vspace{0.08cm}
\begin{minipage}[b]{.46\textwidth}
 \caption{\textbf{Comparison of EXAFS spectra:} \textbf{(a)} Pd-K-edge, \textbf{(b)} Ni-K-edge and \textbf{(c)} P-K-edge EXAFS spectra of \bmg. The experimental data~\cite{Kumar} is shown by red dots, FEAR300 is represented by the blue line and MQ200 is shown by the black line. The \textbf{Inset} in each figure shows the EXAFS spectrum i.e.  \big{(}$k^2 \chi(k)$\big{)}. The FEAR300 model shows a better correlation with the experimental EXAFS spectra. The corresponding Fourier transformations of FEAR300 and MQ200 (blue label) is plotted
alongside experimental values~\cite{Kumar}(red label) and it is observed FEAR300 model is qualitatively in better agreement with the experiment.  }\label{FIG:EXAFS}
\end{minipage}%\vspace{0.3cm}
\end{figure*}

\section{RESULTS AND DISCUSSION}

\subsection{Structural Properties}
Structurally, atoms in \bmg form a densely packed structure with a coordination ranging between 6 and 16, (see Fig.\ref{Fig_Coordination}). The average coordination of Pd, Ni and P are  13.17, 12.15 and 8.95 respectively (Table I). In Table II we present details of coordination number of individual atomic species of the  FEAR300 model. For Pd, the coordination number lies in the range 11-16 with 12-fold, 13-fold and 14-fold coordinated at 22.5\%, 34.165\%, 26.665\% respectively. It is also observed that Pd preferably forms bond with Pd-atoms (40\%-62\%). A similar examination for Ni shows that its coordination number varies between 10-15 with 12-fold and 13-fold being  most abundant. From Table II we observe that Ni-Ni bonds are most common followed by Ni-Pd bonds and Ni-P bonds. In case of P, Fig. \ref{Fig_Coordination} shows that the coordination number can vary from 6-11 with 9-fold coordinated being the most common. The P atoms bonds mostly with Ni (49.36\%) and Pd (41.49\%) while P-P bonding is less observed. The higher fraction of P-Ni bonds are consistent with previous studies~\cite{Kumar,Guan}. The coordination plot in  Fig. \ref{Fig_Coordination} shows a striking similarity between FEAR300 and MQ200 models.

The nature and degree of short range order in metallic glasses is correlated with topology and is quite 
sensitive to small changes in composition~\cite{Miracle1,Miracle2}. Short-range properties have 
a direct impact on glass formation and its stability~\cite{Kelton}. In Fig.~\ref{Fig_Coordination} we show short range properties of \bmg with the plots of total radial distribution function (RDF) and partial radial distribution function. The first minimum in the total RDF occurs at 3.4 $\mathring{A}$ for both the 
MQ200 and FEAR300 models, consistent with the experimental RDF. From partials we further observe that the P-P bonding are less common and form around 4$\mathring{A}$. While, Ni-P bonds peak around 2.8 $\mathring{A}$, Pd-P bonds at 2.9 $\mathring{A}$. This observation is also highlighted by the average coordination ( see Table I)  which are consistent with previous finding~\cite{Kumar}.

We further interrogate the structure by computing the Extended X-ray Absorption Fine Structure (EXAFS) spectra. EXAFS provides valuable first shell information~\cite{Filipponi1}, and is a key structural experiment especially for multicomponent system in which  the existence of several partial pair-correlation functions makes the total pair correlation function far less informative than in an elemental system~\cite{Egami1, Bishal3}.
The EXAFS spectra for both BMG models were calculated using \textit{ab initio} code FEFF9.~\cite{FEFF1} We have used a cluster radius of 5.5 $\mathring{A}$ centered on the absorber atom (Pd, Ni or K, see Fig.~\ref{FIG:EXAFS}). The obtained K-edge spectra of each cluster were averaged over to obtained a final spectrum. Similarly, we obtain Fourier transformation of the EXAFS spectrum by using a Hanning window function with transform range from 2.0 to 12.0 $\mathring{A^{-1}}$ and $dk=0.05$. The Fourier transformation was obtained using the IFEFFIT software.~\cite{IFEFFIT1} We have compared our result with 
the experimental data of \textit{Kumar et. al.}~\cite{Kumar} and our results have a good agreement with the experimental data. Interestingly, our 300 atom model obtained with FEAR has an excellent correlation with the \textit{melt-quench} model and 
the experimental results. The first peak in Fourier transform of EXAFS spectra represents Pd-P and Ni-P peak in Fig.~\ref{FIG:EXAFS}. Similarly, the 
second peak observed $\sim$ 2.5 $\mathring{A}$ represents Pd-Pd(Ni) and Ni-Pd(Ni) in Fig.~\ref{FIG:EXAFS}. 

\subsection{Electronic Properties}

\begin{figure*}[!ht]
\begin{minipage}[b]{.5\textwidth}
  \centering
  \includegraphics[width=0.975\linewidth]{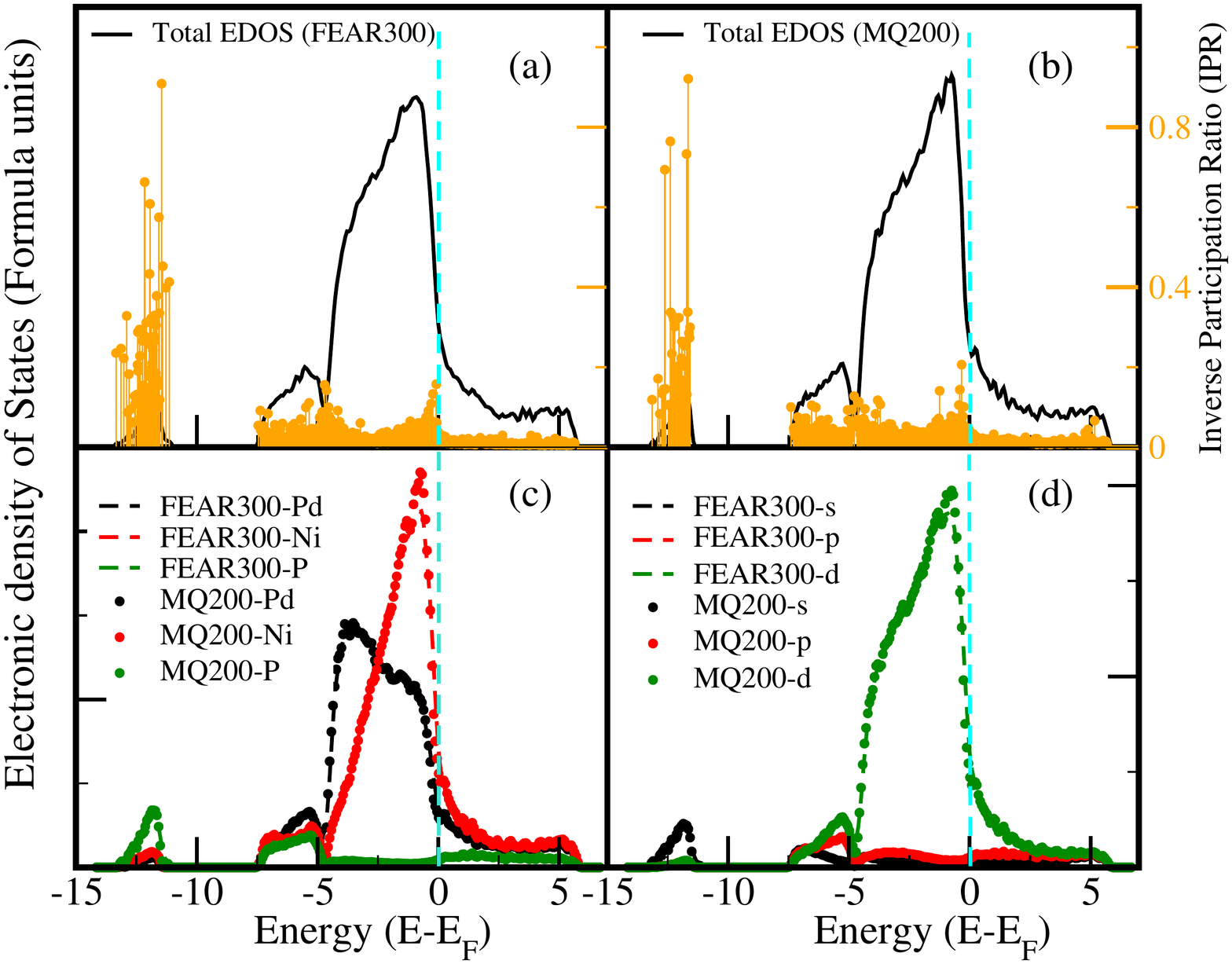}
\end{minipage}%\hspace{0.2cm}
\begin{minipage}[b]{.50\textwidth}
  \centering
  \includegraphics[width=0.975\columnwidth]{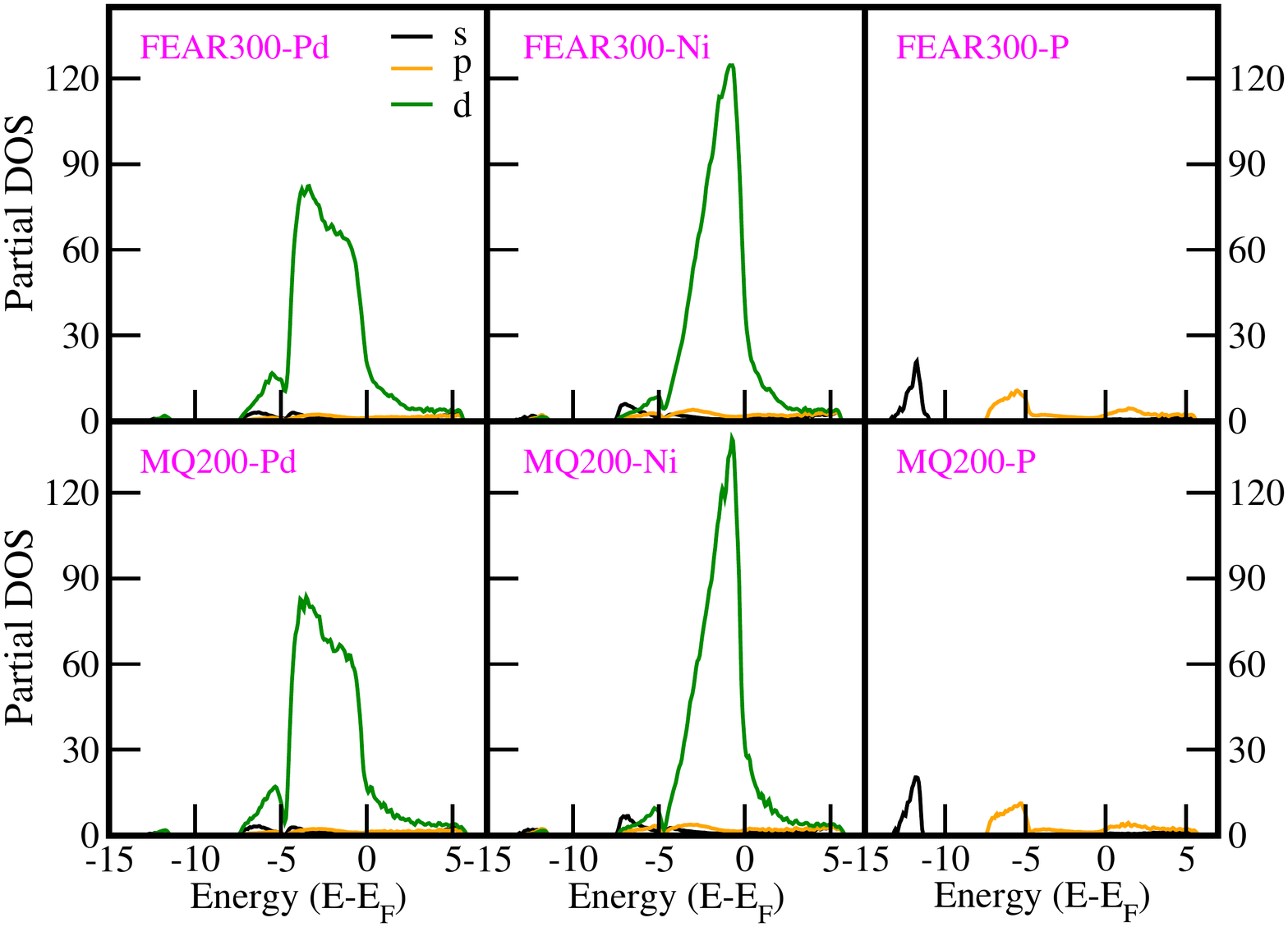}
\end{minipage}%\vspace{0.2cm}
\caption{\textbf{Electronic density of states (EDOS): }\textbf{(Left panel)} \textbf{(a, b)} Total EDOS of FEAR300 and MQ200 models. The total EDOS obtained from VASP was normalized by dividing  3 and 2 respectively for FEAR300 and MQ200 to form formula units of \bmg  with 100 atoms. We indicate localization of electronic states by plotting IPR (yellow drop lines). The electronic states near Fermi level are quite extended. \textbf{(c)} Species projected EDOS of FEAR300 and MQ200 model. Ni and Pd seems to dominate the EDOS contribution. \textbf{(d)} Orbital projected EDOS of FEAR300 and MQ200 model. The d-orbital clearly dominates the EDOS. \textbf{(Right panel)} Plot of s,p,d orbitals projected into respective species. The d-orbitals of Pd and Ni significantly contribute to the EDOS.}\label{fig:Parital_DoS}
\end{figure*}

The electronic properties of the two models were studied by evaluating total and partial 
electronic density of states, and associated localization. The electronic density of states (EDOS) is shown in Fig.~\ref{fig:Parital_DoS}. The Fermi level has been shifted to zero as shown by the dashed horizontal line.  Significant contributions to the total EDOS arise from the Ni and Pd atoms while the P-atoms contribute to the total EDOS at energies deep into valance band. The energy range -7.5 eV to 5.0 eV mainly arises from hybridization of p-orbitals of P with d-orbitals of Pd and Ni (bonding states)~\cite{Kimura}. This is further highlighted in Fig.~\ref{fig:Parital_DoS}. Similarly, s- and p-components in the EDOS of Pd and Ni are small compared to the d-component in agreement with previous results~\cite{Kumar,Guan}. The hybridization of p-orbitals of P with d-orbital of Pd and Ni above the Fermi level gives rise to  antibonding states~\cite{Kimura,Kumar}. The bonding between P and Pd/Ni is reported be of covalent type for a similar composition~\cite{Qing1}.

To further highlight the nature of electronic states we define the 
electronic inverse participation ratio (IPR) as:

\begin{equation}
 \mathcal{I}(\psi_n)= \frac{\sum {a_{i}}^4 }{\big (\sum {a_{i}}^2\big)^2}
\end{equation}

Here, $a_i$ are the components of eigenvector projected onto atomic s, p, and d states as obtained from VASP. The IPR of electronic states is a measure of localization. A localized state would have an IPR value very high (ideally equal to $\mathcal{I}=1$) while a completely  extended state has a value of 1/\textit{N} i.e. evenly distributed over \textit{N} atoms. The IPR is small near the Fermi level, implying extended states indicating of course that \bmg  structure is conducting.

\begin{figure*}[!ht]
\begin{minipage}[b]{.5\textwidth}
  \centering
  \includegraphics[width=0.975\linewidth]{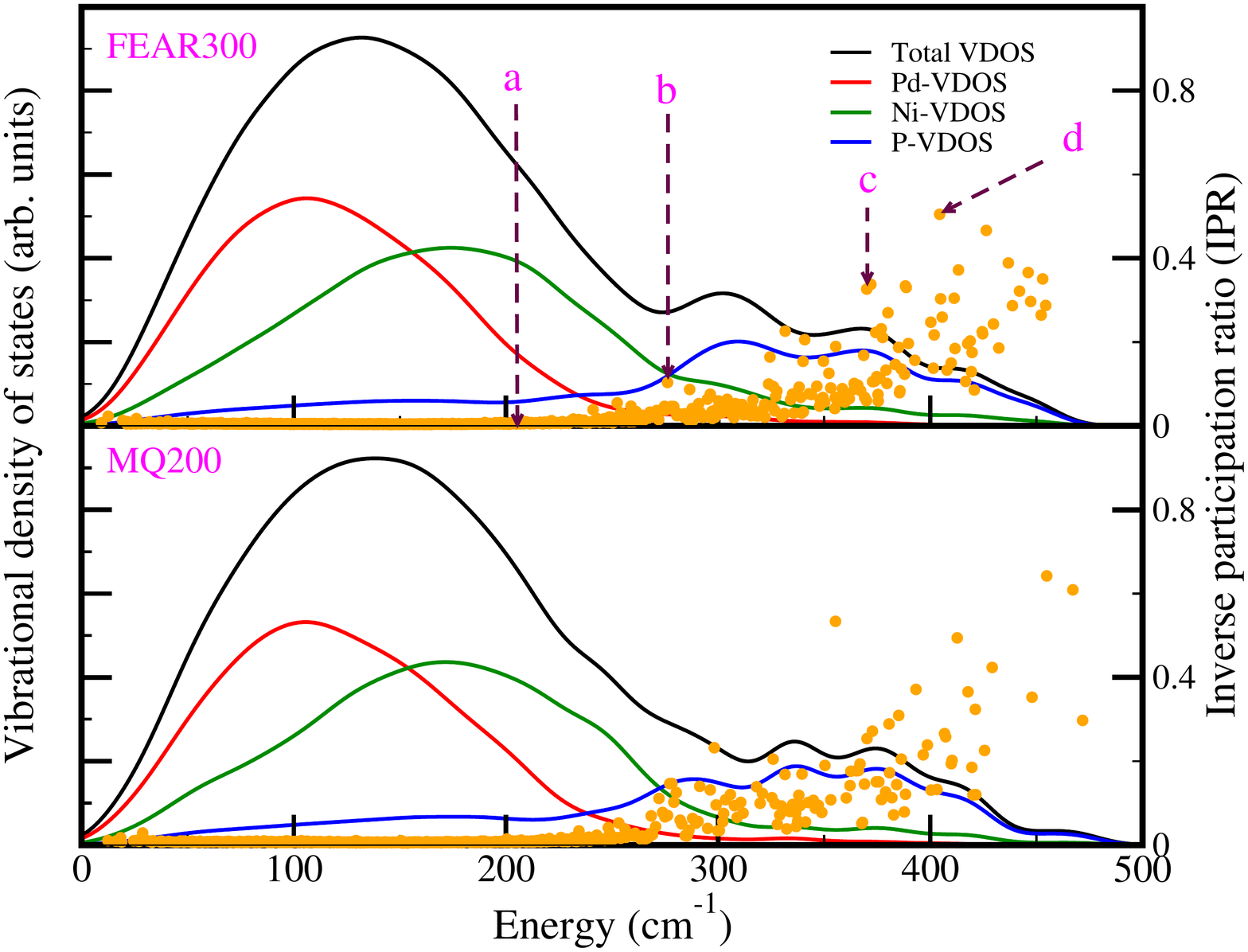}
  \label{fig:fig1023114}
\end{minipage}%\hspace{0.2cm}
\begin{minipage}[b]{.5\textwidth}
  \centering
  \includegraphics[width=0.975\linewidth]{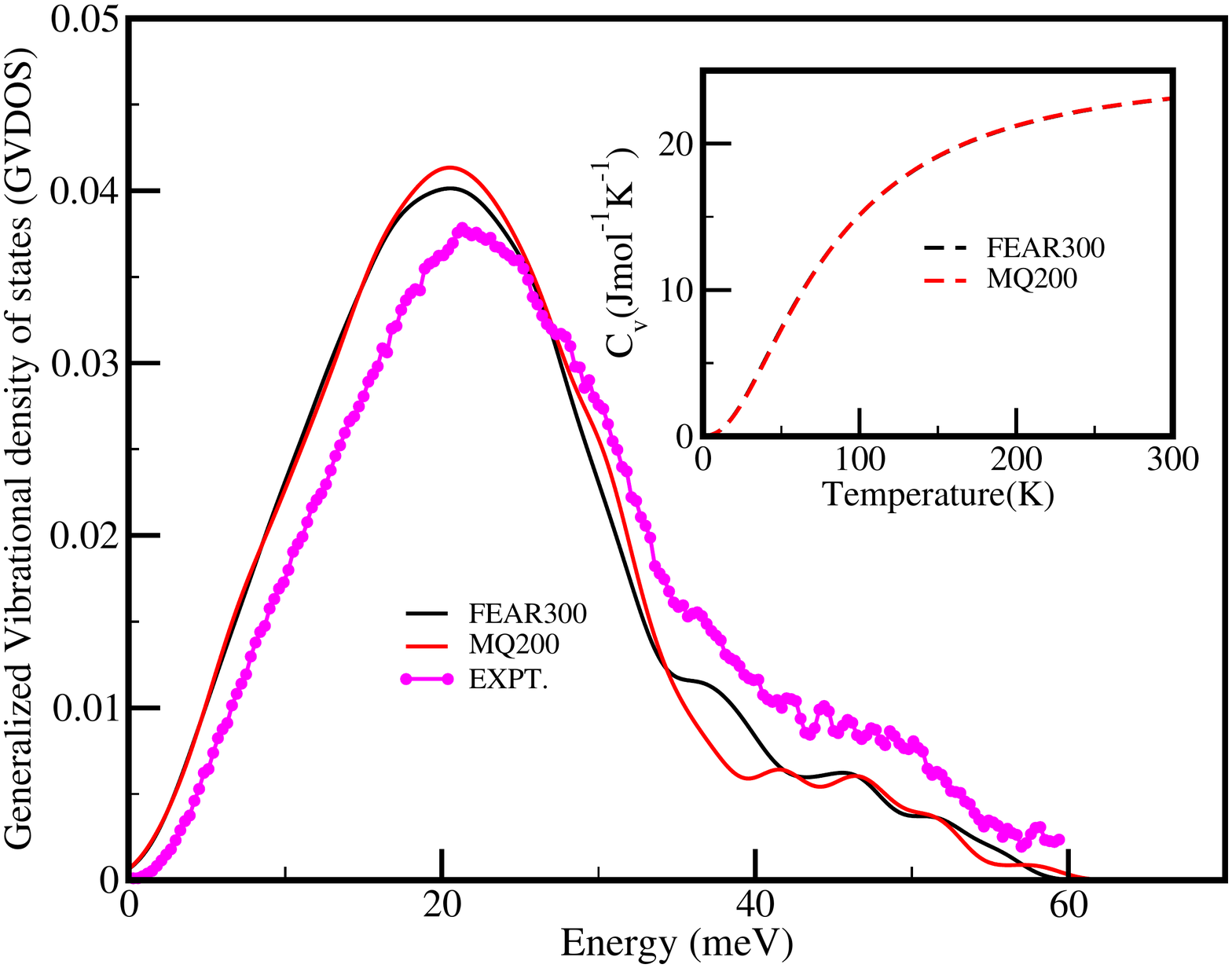}
  \label{fig:fig1023115}
\end{minipage}%\hspace{0.2cm}
\caption{\textbf{Vibrational density of states (VDOS):} (\textbf{Left panel}) We show VDOS plots obtained for FEAR300 and MQ200 models. The total VDOS (black line) shows a maximum peak around $\sim 150 cm^{-1} \approx 19 meV$. The decomposition of VDOS into individual species contribution highlights that $Pd$-VDOS (red line) and $Ni$-VDOS (green line) contribute mostly to this maximum peak. The contribution of $P$-VDOS (blue line)  is mostly at high frequency range. We also show localization of vibrational modes by plotting vibrational inverse participation ratio (VIPR). The VIPR plot (yellow dots) shows that vibrational motion is mostly extended upto $\sim 250 cm^{-1}$ and localized modes appear at higher frequencies. Frequencies occurring at \textbf{(a,b,c,d)} will be explained in Fig.~\ref{Fig:Jmol}.(\textbf{Right panel}) We compare our vibrational spectrum FEAR300 and MQ200 directly with the inelastic neutron scattering results of \textit{J. B. Suck}~\cite{Suck2002}. The generalized VDOS (GVDOS) is obtained via. Equation~\ref{Eq:GVDOS}. Both the models shown reasonable agreement with the experiment. (\textbf{Inset}) We plot the specific heat ($C_v(T)$) obtained from the harmonic approximation (Equation~\ref{Eq:CV}). The specific heat linearly increases with the increase in temperature and follows Dulong and Petit limit at the higher temperatures. }\label{Fig:GVDOS}
\end{figure*}

\subsection{Vibrational Properties}

\begin{figure}[!ht]
  \centering
   \includegraphics[width=9.0 cm,]{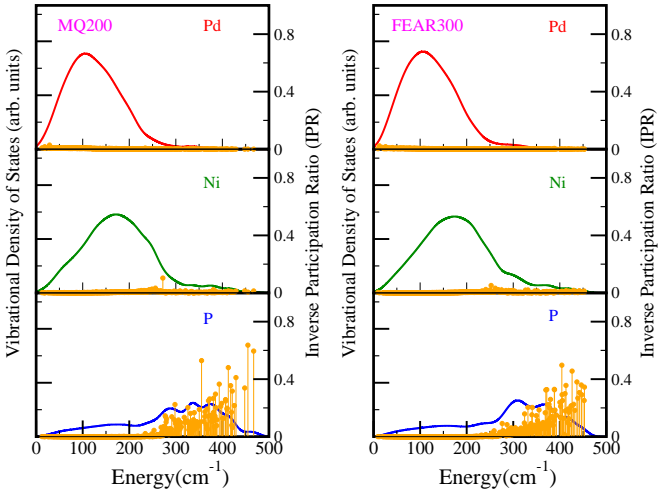}
  \caption{\textbf{Species projected VDOS and VIPR:} We decompose the VDOS and VIPR to analyze the contribution of each individual species. It is seen that P-atom contributes to almost all the localization of vibrational modes that occur at higher frequency range.}
  \label{Fig:ProjectedVIPR}
\end{figure}

\begin{figure*}[htp]
\begin{minipage}[b]{.45\textwidth}
  \centering
  \captionsetup{justification=centering,singlelinecheck=false}
  \caption*{\large \textbf{(a) IPR = 0.0045}}
  \includegraphics[width=0.96\linewidth]{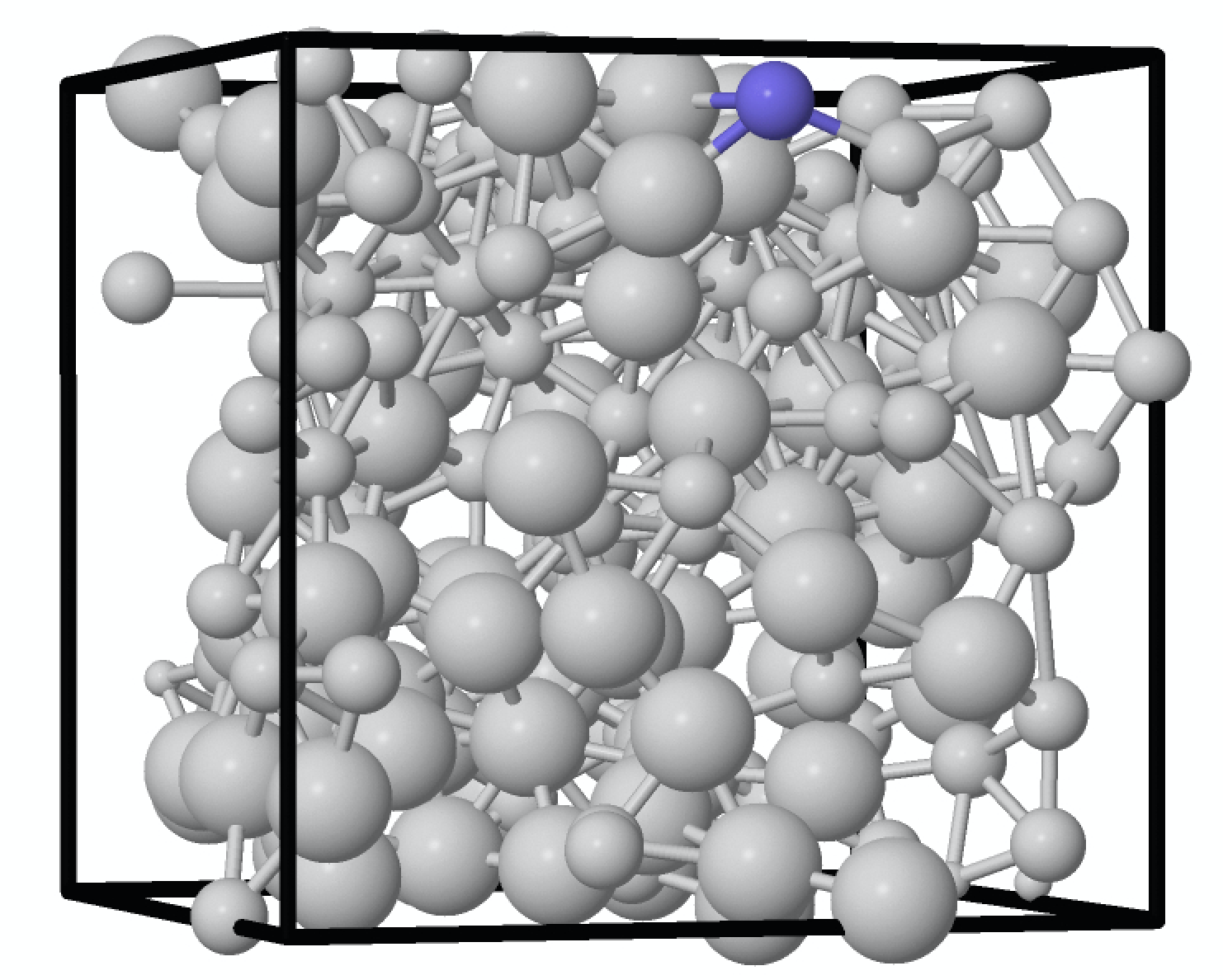}
\end{minipage}\vspace{0.45cm}
\begin{minipage}[b]{.45\textwidth}
\vspace{0.15cm}
 \centering
 \captionsetup{justification=centering,singlelinecheck=false}
  \caption*{\large \textbf{(b) IPR = 0.1037}}
  \includegraphics[width=0.84\linewidth]{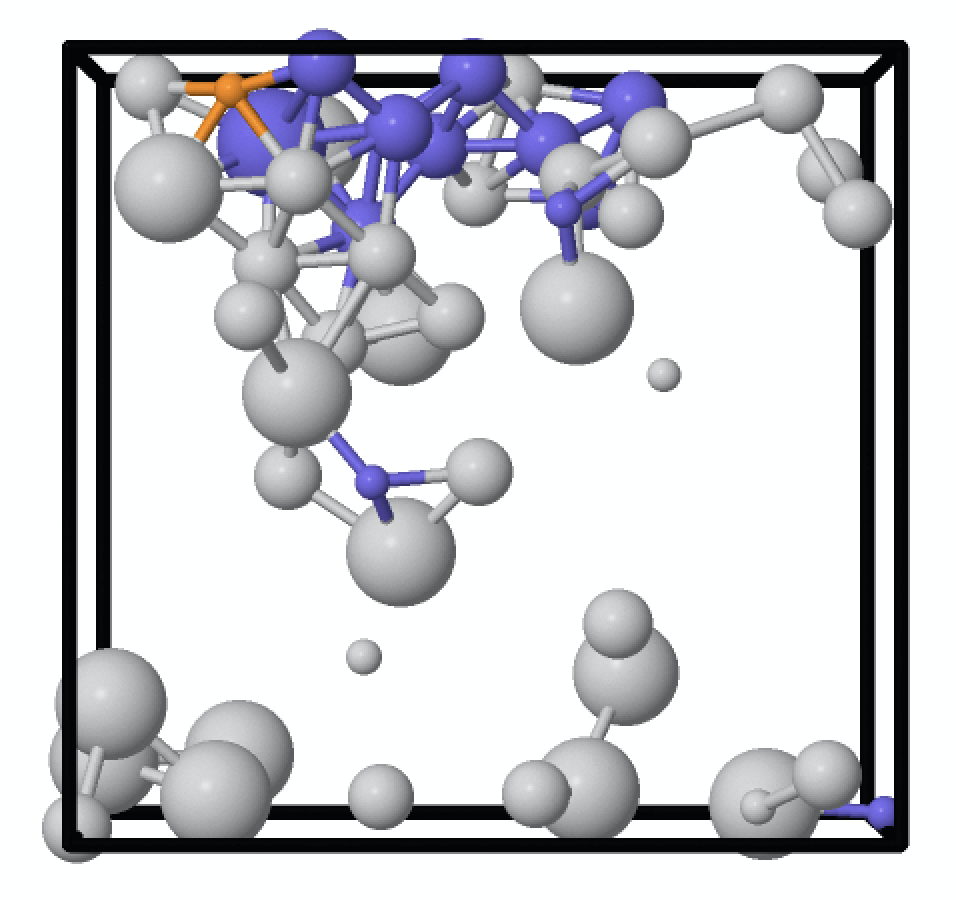}
\end{minipage}\vspace{0.10cm}
\begin{minipage}[b]{.45\textwidth}
 \centering
 \captionsetup{justification=centering,singlelinecheck=false}
  \caption*{\large \textbf{(c) IPR = 0.3265}}
  \includegraphics[width=0.98\linewidth]{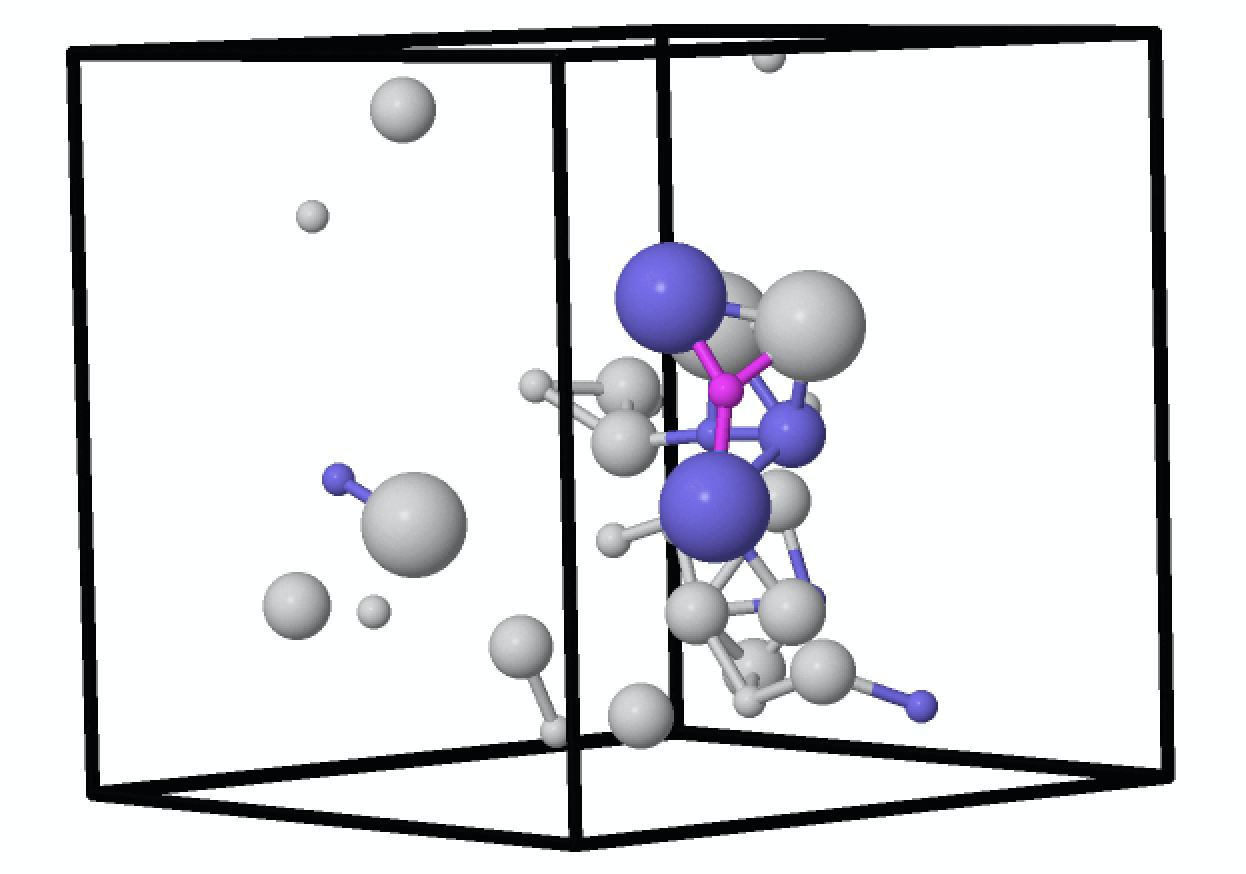}
\end{minipage}\vspace{0.1cm}
\begin{minipage}[b]{.45\textwidth}
\vspace{0.20cm}
\captionsetup{justification=centering,singlelinecheck=false}
  \caption*{\large \textbf{(d) IPR = 0.5050}}
 \centering
  \includegraphics[width=0.75\linewidth]{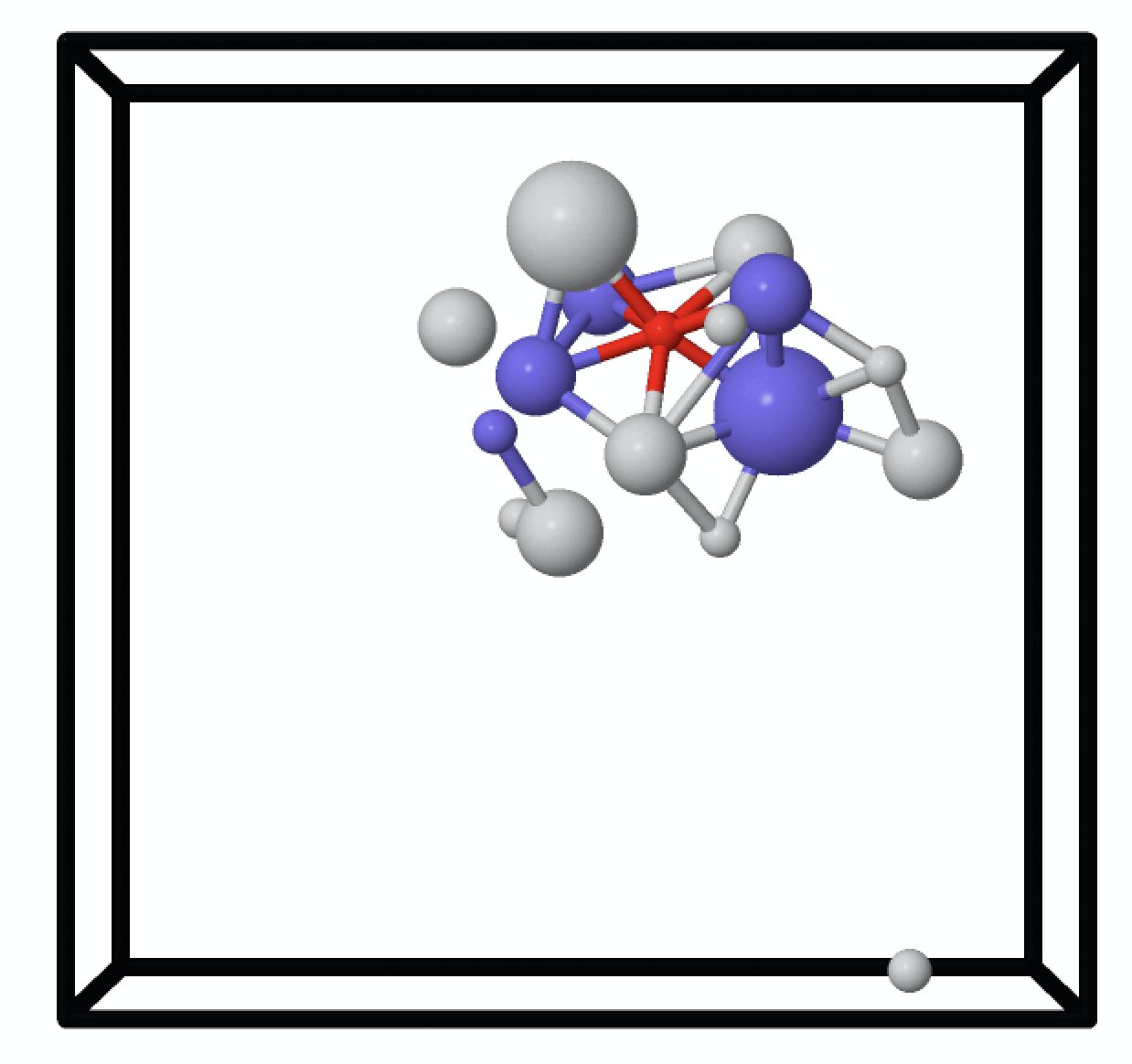}
\end{minipage}\vspace{0.1cm}
   \caption{\textbf{Localized-delocalized vibrational transition in FEAR300:}  The atoms active in particular vibrational mode. Each atom labeled with different colors reflect the fraction of total vibration at that frequency. 
   Color scheme:  grey ($< 1.0\% $\hspace{0.1cm} and $\geq 0.1\%$),\hspace{0.1cm} blue ($< 15.0\% $\hspace{0.1cm} and $\geq 1.0\%$),\hspace{0.1cm}
  green ($< 30.0\% $\hspace{0.1cm} and $\geq 15.0\%$),\hspace{0.1cm}  orange ($< 45.0\% $\hspace{0.1cm} and $\geq 30.0\%$),\hspace{0.1cm}
  magenta ($< 60.0\% $\hspace{0.1cm} and $\geq 45.0\%$),\hspace{0.1cm} red ($< 75.0\% $\hspace{0.1cm} and $\geq 60.0\%$)\hspace{0.1cm} and 
  brown( $\geq 75.0\%$)\hspace{0.1cm}. The atom size is representative of respective atoms ($Pd$ (largest) to $P$ (smallest)). The labels (a,b,c,d) on top of each figure is highlighted in Fig.~\ref{Fig:GVDOS} (left panel). \textbf{(a)} $\omega=204.941 \hspace{0.05cm}cm^{-1}$ and IPR = 0.0045, 
  \textbf{(b)} $\omega=276.162 \hspace{0.05cm}cm^{-1}$ and IPR = 0.1037, \textbf{(c)} $\omega=370.04 \hspace{0.05cm}cm^{-1}$ and IPR = 0.3265 and \textbf{(d)} $\omega=404.405 \hspace{0.05cm}cm^{-1}$ and IPR = 0.5050. In figure (c,d) we can observe that phosphorous atom has $\sim 50\%$ or 
  more of the total vibration.}\label{Fig:Jmol}
\end{figure*}

To our knowledge, this paper is the first attempt to thoroughly 
study the lattice dynamics of \bmg.
Vibrational properties provide special insight of the local bonding in any material, and thanks especially to inelastic neutron scattering, are readily compared to computations of the density of states~\cite{Thorpe}. Vibrational and thermal properties offer a key test to validate (or invalidate!) computer models. Typically, vibrational propertes in amorphous material has been mostly studied using: (a) Fourier transformation of two-point velocity auto-correlation function and, (b) harmonic approximation or frozen phonon calculation. Both methods have their own advantages and limitations.  Owing to system size and computational expense, we have used the harmonic approximation to study vibrational properties of \bmg. Accurate computations of the GVDOS are difficult, requiring accurate interatomic potentials, the ``right" topology of the models and large systems to minimize size artifacts.

To determine the vibrations, we first relax both models to attain zero pressure. This relaxation resulted in a slight increase of volume  ($\sim 4 \%$), no significant network topology changes and non-orthogonal lattice vectors for the supercell. We then compute the Hessian by displacing each atom by 0.015 $\mathring{A}$ in 6-directions ($\pm x$,$\pm y$,$\pm z$). We form and diagonalize the dynamical matrix at the zone center, and compute the density of states by Gaussian broadening the eigenvalues from Equation~\ref{Eq:VDOS} (see details ~\cite{Bhattarai1,Bhattarai2}). The first three frequencies are very close to zero,  and arise from rigid supercell translations, and we have therefore these in what follows. The vibrational density of states (VDOS) is,

\begin{equation}
g(\omega)=\frac{1}{3N} \sum_{i=1}^{3N} \delta(\omega-\omega_i)
 \label{Eq:VDOS}
\end{equation}
Where, $\omega_n$  represent eigenfrequencies of normal modes and ``\textit{N}" is the number of atoms.

Similarly, we can evaluate the species projected VDOS as,
\begin{equation}
g_{\alpha}(\omega)=\frac{1}{3N} \sum_{i=1} ^{N_{\alpha}} \sum_{n} \lvert e_{i}^{n}\lvert^{2} \delta(\omega-\omega_n)
\label{Eq:VDOSPartial}
\end{equation}
Here, $  e_{i}^{n}$ are the eigenvectors of the normal modes and $N_{\alpha}$ is total number of atom for $\alpha$ species.

Experimentally, by using inelastic neutron scattering, the vibrational density of states is directly evaluated in terms of generalized vibrational density of states (GVDOS) $G(\omega)$. The GVDOS is defined as~\cite{Suck2002},

\begin{equation}
G(\omega)=\frac{\sum w_{i} g_{i}(\omega)}{\sum w_i } 
 \label{Eq:GVDOS}
\end{equation}

Here, $w_{i}=\frac{e^{-2W_i}c_i\sigma_{i,sc}}{M_i}$ is a weighting factor which depends upon the Debye-Waller factors, mass of the species and so on. We have used taken  $w_i$ as 0.768 (Ni), 0.102 (Pd) and 0.13 (P) as used in the experiment.~\cite{Suck2002} . We compare our results for FEAR300 and MQ200 in Fig.~\ref{Fig:GVDOS}. Our results shows reasonable agreement with the inelastic neutron scattering result. The VDOS of \bmg shows a signature peak value around $\sim 150 cm^{-1} \approx 19 meV$. This peak is mostly contributed by Pd-atom and Ni-atom vibrations. The VDOS of Pd-atom peaks around $\sim 100 \hspace{0.05cm} cm^{-1} $ and Ni-atom VDOS peaks around $\sim 180 \hspace{0.05 cm}cm^{-1} $.  There is no significant contribution of P-atom in the lower frequency regime.  The FEAR300 and MQ200 models produce quite similar spectra. We also provide a direct comparison of GVDOS with experimental results in Fig.~\ref{Fig:GVDOS}. The experimental GVDOS and both of our models show reasonable agreement.

After obtaining $g(\omega)$ the vibrational specific heat can be evaluated by using the following relation,

\begin{equation}
 C(T)= 3R\int_0^{E_{max}}\Bigg(\frac{E}{k_{B}T}\Bigg)^2 \frac{e^{E/k_BT}}{\Big(e^{E/k_BT}-1\Big)^2}g(E)dE
 \label{Eq:CV}
\end{equation}

Here, $g(E)$ in normalized to unity~\cite{Maradudin1}. \vspace{0.1cm}

The evaluation of the vibrational specific heat within the harmonic approximation is straightforward with the knowledge of vibrational density of states i.e. $g(\omega)$. We compute the specific heat as shown in Equation~\ref{Eq:CV}.  Our plot obtained for specific heat for the two models is shown in Fig.~\ref{Fig:GVDOS} \textbf(inset). The specific heat for both models increases almost linearly with the temperature (T) before it starts saturating to Dulong and Petit limit for higher temperature. However, we do not observe any boson peak seen in some metallic glasses in our models~\cite{Yong1}. This is perhaps unsurprising for small models~\cite{Crespo}.

\subsubsection{Vibrational Localization}

While the density of states may be accurately inferred from experiments, the structure and extent of the associated vibrational eigenvectors is not directly observable.  To further work out the nature of the vibrations in \bmg, we look into the localization of vibrational eigenstates by calculating vibrational inverse participation ratio (VIPR). Similar to electronic IPR, VIPR can be readily calculated from the eigenvectors as shown in 
Equation~\ref{Eq:VIPR}.

\begin{equation}
\mathcal{I}(u^j_i)=\frac{\sum_{i=1} ^N |u^j_i|^4 }{\big(\sum_{i=1} ^N |u^j_i|^2\big )^2}
\label{Eq:VIPR}
\end{equation}\vspace{0.1cm}
Where, ($u^j_i$) is normalized eigenvector of $j^{th}$ mode.

Small values of VIPR signify evenly distributed vibration among the atoms while higher value imply that only a few atoms contribute to the total vibration at that particular eigenfrequency. We have plotted the total VIPR in Fig.~\ref{Fig:GVDOS}. The vibration up to $\sim 250 \hspace{0.05cm}cm^{-1}$ are completely extended while vibrations start to localized after $ 250 \hspace{0.05cm}cm^{-1}$. The evolution of localization is fairly smooth and spans a very broad range of localization from purely extended to compactly localized.  This is seen for both FEAR300 and MQ200 models. To further investigate the nature of localization occurring at higher frequencies we evaluate species projected VIPR. This projection of VIPR is evaluated such that the contribution of each individual atom sums up to the total VIPR as shown in Equation~\ref{Eq:VIPRProjected}.

\begin{equation}
\mathcal{I}_{total}(u^j_i)=\mathcal{I}_{Pd}(u^j_i)+ \mathcal{I}_{Ni}(u^j_i)+ \mathcal{I}_{P}(u^j_i)
\label{Eq:VIPRProjected}
\end{equation}\vspace{0.1cm}

This decomposition of VIPR shows that the localization occurring at higher frequencies is exclusively due to P-atoms. Their role in high frequency oscillations is obvious to some degree since the P atoms are of course lighter than the Pd or Ni. However, the concentration of P is high -- 20\%, so that one would imagine that there would be ``banding" between the P atoms distributed through the cell. Thus we make the simple point that unlike a system like H in Si (which possesses a drastic mass difference and a small H concentration), it is not obvious that there should be strongly localized P vibrational modes in our system, even at high frequency. We project vibrational contribution at particular frequencies onto their corresponding atoms. This assignment is done by including all the atoms that 
 participate to contribute 90 \% of vibrations at that frequency. The color scheme (see caption~\ref{Fig:Jmol}) represents different percentage of vibrations of atoms at that frequency and the size of atom is representative of our composition ($Pd$ (largest) to $P$ (smallest)). We visualize these modes starting from extended to localized modes at different frequencies (see a,b,c,d in Fig.~\ref{Fig:GVDOS}). In Fig.~\ref{Fig:Jmol} (a) we show projection of vibration for  IPR ($\mathcal{I}$) value of 0.0045. This is an extended vibrational mode and from Fig.~\ref{Fig:Jmol} (a) we see that almost all the atoms are in motion with most of them having vibration ranging between $< 1.0\% $\hspace{0.1cm} and $\geq 0.1\%$\hspace{0.1cm}. In Fig.~\ref{Fig:Jmol} (b) with $\mathcal{I}$ = 0.1037, we start to see some blue color atoms which indicate few $Ni$ and $Pd$ have vibrations ranging  between $< 15.0\% $\hspace{0.1cm} and $\geq 1.0\%$.\hspace{0.1cm}
As we move towards more localized states (Fig.~\ref{Fig:Jmol} (c) and (d)) with $\mathcal{I}$ values of 0.3265 and 0.5050 we see single $P$ atom is contributing $45- 75\%$ of the total vibrations. These localized modes occur at higher frequency where stretching modes mostly dominate with few bending type of modes. The red phosphorus atom in  Fig.~\ref{Fig:Jmol} (d) contains more than 60 \% of the total vibration at that frequency.

Vibrational localization plays a central role in the thermal conductivity of materials, analogous to the situation for electrons. If we imagine ``tuning" the frequency from $ 250 \hspace{0.05cm}cm^{-1}$ to the high energy end of the spectrum in Fig. \ref{Fig:Jmol}, the P atoms fully participate at the beginning but become confined rattlers at the higher frequencies. Heat transport is essentially limited to normal modes below ca. 400 cm$^{-1}$. It would be quite interesting to apply vibrational hole-burning experiments to these systems. Hole lifetimes would be closely related to the localization that we indicate here. Further out on a limb, these observations suggest that such systems could be worth exploring for thermoelectric applications (the ``electron-crystal,  phonon-glass" picture). We do not suggest that this composition is well suited for this, but might motivate new directions of experimental and modeling inquiry.

\section{ Conclusions} 

We discover an interesting new localized to extended transition in the vibrational states and show that the high energy, localized modes are associated with trapped P.  We have found that AIFEAR is a promising method to model several complex systems. \textit{Ab initio} shows similar character to the conventional \textit{ab initio} MD model despite requiring fewer force calls. The structural, electronic and vibrational properties of FEAR model are in good agreement with observed experiments and previous literature. The EXAFS spectrum further highlights structural similarity of FEAR model with experiment.  The structural analysis of \bmg highlights that the network is dominated by $Ni-Ni, Pd-Pd$\hspace{0.1cm} and $Pd/Ni-P$ bonds. The rarity of $P-P$ bonds helps to explain highly localized vibrations of $P$ atoms with up to $45- 75\%$ of vibrational motion in that mode. The electronic signatures indicate this material exhibits fairly extended electronic states near the Fermi level.  We have established an accurate \textit{ab initio} model for the \bmg composition and we hope that it will serve as a benchmark for future calculation of complex metallic glasses.

\section{Acknowledgment} 
The authors are grateful to the NSF under grant numbers DMR 1506836, DMR 1507670.
We acknowledge computing time provided by BRIDGES at the Pittsburgh supercomputer center (Allocation ID: DMR180031P) under the Extreme Science and Engineering Discovery Environment (XSEDE) supported by National Science Foundation grant number ACI-1548562. We also thank NVIDIA Corporation for donating a Tesla K40 GPU which was also used in our calculations. \\

\bibliography{sample}

\end{document}